\documentstyle[aps]{revtex}

\author{H. Falomir
\\
Departamento de F\'{\i}sica\\
Facultad de Ciencias Exactas,\\
Universidad Nacional de La Plata, Argentina
}

\title{Condiciones de contorno globales para el
operador de Dirac.\thanks{Talk given at
SOCHIFI X, November 1996, Valpara\'{\i}so, Chile. } }

\date{ noviembre de 1996}

\def\dfrac#1#2{{\displaystyle {#1 \over #2}}}
\def\QATOP#1#2{{#1 \atop #2}}
\def\QDATOP#1#2{{\displaystyle {#1 \atop #2}}}

\def\dint{\displaystyle \int }
\def\k{\mbox{\large $\kappa$}}
\def\ka{\mbox{$\kappa$}}
\def\nn{\nonumber}
\def\be{\begin{equation}}
\def\ee{\end{equation}}

\begin{document}

\maketitle

\begin{abstract}

\end{abstract}

\section*{Determinantes funcionales en teor\'{\i}a cu\'antica de campos}

Las correcciones al orden de {\it 1-loop} en teor\'{\i}a cu\'antica de campos y
sus aplicaciones  a la mec\'anica estad\'{\i}stica conducen naturalmente al
c\'alculo de {\it determinantes} de operadores diferenciales. En efecto, en la
formulaci\'on basada en la integral funcional (en un  espacio - tiempo
Eucl\'{\i}deo $\nu$-dimensional), se trata de evaluar  integrales {\it
Gaussianas} de la forma
\be
\int d\mu[\psi] \, e^{- \int \psi^\dagger D \psi \, d^\nu x } \sim \left({\rm
Det} D\right)^\alpha,
\ee
donde $\alpha = -1/2, -1, +1$, dependiendo de que el campo $\psi$ sea real,
complejo o de {\it Grassmann}, respectivamente.

Esa  expresi\'on es entendida como la generalizaci\'on del resultado que se
obtiene para una integral m\'ultiple sobre cada tipo de variable, para el caso
en que $D$ es simplemente una matriz.

Ahora bien, como $D$ (operador diferencial que aparece en la parte cuadr\'atica
en $\psi$ de la acci\'on) es no acotado, su determinante no puede ser definido
como el producto (divergente) de sus autovalores. Entonces, es necesario
adoptar un esquema de {\it regularizaci\'on} que permita dar un sentido a la
integral funcional.

Por otra parte, cuando los campos est\'an definidos sobre variedades con
borde, las  condiciones de contorno a las que est\'an sometidos
tienen una influencia  relevante sobre el comportamiento de esos determinantes.
 El
estudio de efectos de borde en este
contexto\cite{r0,r1,r2,r3,r4,r5,r6,r7,r8,r9} resulta de inter\'es en
teor\'{\i}a cu\'antica de campos,  por ejemplo, para la descripci\'on de
modelos efectivos para las
interaciones fuertes, la cosmolog\'\i a cu\'antica, o incluso
aplicaciones a la descripci\'on de sistemas de inter\'es en materia condensada.

\subsection*{Operadores el\'{\i}pticos}

Si $D$ es un operador diferencial lineal de orden $\omega$ en una regi\'on
$\Omega$ de ${\rm \bf R}^\nu$,
\be
D=\sum_{|\alpha| \leq \omega} a_\alpha (x) \, (- i \partial_x)^{\alpha}
\ee
(donde $\alpha=(\alpha_1, ..., \alpha_\nu)$, $|\alpha|=\alpha_1 +... +
\alpha_\nu$, y supondremos que los coeficientes $a_\alpha (x) \in C^\infty$),
su {\it s\'{\i}mbolo} en $x\in \Omega$ es un polinomio de grado $\omega$,
definido por
\be
\sigma (D)(x,\xi) = \sum_{|\alpha| \leq \omega} a_\alpha (x) \,  \xi^{\alpha},
\ee
donde $\xi \in {\rm \bf R}^\nu$. El {\it s\'{\i}mbolo principal} es la parte
homog\'enea de grado $\omega$ del s\'{\i}mbolo,
\be
\sigma_\omega (D)(x,\xi) = \sum_{|\alpha| = \omega} a_\alpha (x) \,
\xi^{\alpha}.
\ee

Sea una funci\'on suave $f(x) \in {\cal S}$, y sea $\hat f (\xi)$ su
transformada de Fourier. La acci\'on de $D$ sobre $f$ puede ser expresada como
\be
D f(x) = \frac 1 {(2\pi)^\nu} \int e^{i x \cdot \xi} \sigma (D)(x,\xi) \hat f
(\xi) \, d^\nu \xi.
\label{pseudo}
\ee

M\'as generalmente, dada una funci\'on $\sigma (D)(x,\xi)$ (no es
necesariamente un polinomio en $\xi$), tal que
\be
\left| \partial_{\xi}^{\alpha} \partial_{x}^{\beta}\sigma (D)(x,\xi) \right|
\leq C_{\alpha,\beta} \left(1+ | \xi |\right)^{\omega - \alpha},
\ee
para ciertas constantes $C_{\alpha,\beta}$, y con $\omega$ no necesariamente
entero positivo, la ecuaci\'on (\ref{pseudo}) define un operador {\it
pseudo-diferencial} de orden $\omega$.

Un operador $D$ se dice {\it el\'{\i}ptico}  en $x$ si
\be
\sigma_\omega (D)(x,\xi) \neq 0 \quad \forall \, \xi \neq 0.
\ee

Si $D$ es el\'{\i}ptico en una regi\'on {\it compacta} $\Omega$ entonces, para
$ \left| \xi \right|>0$,
\be
\left| \sigma_\omega (D)(x,\xi) \right| \geq const. \, \left| \xi
\right|^\omega >0, \quad
\forall  \, x\in \Omega,
\ee
ya que ambos miembros son homog\'eneos de grado $\omega$, y $a_\alpha (x) \in
C^\infty$.

Por ejemplo, en ${\rm \bf R}^2$, el operador $D=-i( \partial_1 + i \partial_2)$
es el\'{\i}ptico, ya que $\xi_1 + i \xi_2 = 0 \Rightarrow \xi=0$. Lo mismo vale
para el Laplaciano, $\nabla=(\partial_1)^2+(\partial_2)^2$.

\subsection*{Determinantes funcionales de operadores el\'{\i}pticos en
variedades sin borde}

Dado un operador el\'{\i}ptico {\it invertible} de orden $\omega$, $D$,
definido sobre una variedad compacta {\it sin borde} $\Omega$, de dimensi\'on
$\nu$, puede definirse un determinante regularizado empleando las potencias
complejas del operador, introducidas por Seeley\cite{Seeleysb}.

Para ello se construye la funci\'on
\be
\zeta_D (s) =Tr\{D^{-s}\} =  \sum_j (\phi_j , D^{-s} \phi_j),
\label{zetasb}
\ee
donde las $\{\phi_j \}$ forman un conjunto ortonormal completo.

Por ejemplo, para un operador autoadjunto, y tomando la base de sus
autofunciones, $D\,\phi_j = \lambda_j \phi_j$, se tiene
\be
\zeta_D (s) = \sum_j \lambda_j^{-s}.
\label{zeta-auto}
\ee
Las series (\ref{zetasb},\ref{zeta-auto}) convergen para $\Re(s)>\nu/\omega$, y
pueden ser extendidas anal\'{\i}\-ti\-ca\-men\-te en el plano $s$ a funciones
meromorfas, que s\'olo tienen polos simples y que, en particular, son regulares
en $s=0$.

En esas condiciones, se define el deteminante como
\be
{\rm Det} \, D =\left. exp \left( - \frac{d \zeta_D (s)}{d s}\right)
\right|_{s=0}.
\label{det-zeta}
\ee
N\'otese que, para una matriz autoadjunta, el argumento de la exponencial se
reduce a $ \sum_j \ln \lambda_j$, de modo que (\ref{det-zeta}) representa una
regularizaci\'on del producto de autovalores.

En el caso general, $\zeta_D (s)$ puede ser expresada en t\'erminos de la
extensi\'on meromorfa del n\'ucleo de $D^{-s}$ tomado en la diagonal principal,
\be
\zeta_D (s)=\int d^\nu x \, tr\{ J_{-s}(x,x)\}
\ee
(para $\Re(s)$ suficientemente grande, $J_{-s}(x,y)$ es una funci\'on continua
de sus variables (s,x,y), y anal\'{\i}tica en $s$).

Desde luego que otras regularizaciones son posibles. Pero una ventaja de la
definici\'on  derivada de la funci\'on  $\zeta$ del operador es que es
invariante de gauge.

\subsection*{Problemas el\'{\i}pticos de borde}

En lo que sigue estaremos interesados fundamentalmente en operadores
el\'{\i}pticos  de primer orden ($\omega=1$), definidos en variedades
$\nu$-dimensionales, $M\subset {\rm R}^\nu$, con borde $\partial M$ suave,
\begin{equation}
\label{OP}
D:C^\infty (M,E)\rightarrow C^\infty (M,F),
\end{equation}
donde $E$ y $F$ son fibrados complejos $k$-dimensionales sobre $M$.

En general, tales operadores tienen un {\it subespacio nulo},
\be
{\rm Ker}(D)=\left\{\varphi(x) / D\varphi(x) =0, \, x\in M\right\},
\ee
infinito-dimensional.
 Entonces, para tener un problema bien definido, debe restringirse el tipo de
funciones {\it aceptables}. Eso se hace imponiendo condiciones de contorno  que
excluyen de su dominio de definici\'on a {\it casi todas} las soluciones del
operador, dejando s\'olo un {\it kernel} de dimensi\'on finita.

\bigskip

Una herramienta fundamental para estudiar problemas de borde es el operador de
Calder\'on, $Q$. Se trata de una proyecci\'on (no necesariamente ortogonal)
\be
\label{Q}
Q:L^2(\partial M,E_{/ \partial M})\rightarrow \{(T\varphi \ /\varphi \in {\rm
Ker} (D)\},
\ee
donde $T:C^\infty
(M,E)\rightarrow C^\infty (\partial M,E_{/ \partial M})$ toma los valores de
borde de $\varphi$ (operador {\it traza} para operadores de orden 1), siendo
$E_{/ \partial M}$ la restricci\'on de $E$ al borde.

Como se muestra en\cite{Calderon}, $Q$ es un pseudo-diferencial de orden 0,
cuyo s\'{\i}mbolo principal, $q(x,\xi)$, s\'olo depende del simbolo principal
del operador $D$.

\bigskip

Por ejemplo, consideremos  el operador de Dirac,
\begin{equation}
\label{culo}
D(A)=i\not \! \partial +\not \! \! A=
\sum_{\mu =0}^{\nu -1}\gamma _\mu \left( i \frac \partial {\partial
x_\mu }   + A_\mu\right),
\end{equation}
donde $\{A_\mu ,\ \mu =0,...,\nu -1\}$ es el campo de gauge.
En este caso $k=2^{[\nu /2]}$, dimensi\'on de los espinores de Dirac en $\nu$
dimensiones.

El operador $Q$ puede ser construido a partir de cualquier soluci\'on
fundamental de $D$,  empleando la f\'ormula de Green\cite{fat}. En efecto, sea
$K(x, y)$ tal que
\begin{equation}
D^{\dagger }(A)K^{\dagger }( x, y)=\delta ( x- y), \quad x,y\in M.
\end{equation}
Podemos escribir
\begin{equation}
\label{po}K( x, y)=K_0( x, y)+R( x, y),
\end{equation}
donde $K_0( x, y)$ es la soluci\'on fundamental de $i\!\!\not \! \partial$ que
se anula en el infinito,
\begin{equation}
\label{pot}K_0( x, y)=-\ i\ \frac{\Gamma (\nu /2)}{2\ \pi ^{\nu /2}}%
\ \ \frac{( {\not\! x}- {\not\! y})}{\vert  x- y\vert ^\nu },
\end{equation}
y $\vert R( x, y)\vert \sim {\rm O} (1/ \vert  x- y\vert ^{\nu -2})$.

Si $f(x)$ es una funci\'on suave sobre el borde, a valores en $E_{/ \partial
M}$, entonces
\begin{equation}
\label{poto}
Q f(x)=-i\lim \limits_{ x\rightarrow \partial
M}\int_{\partial M}K( x,y)\ \not \! n\ f(y)\ d\sigma _y,
\end{equation}
donde  $\not \!\! n=\sum_\mu \gamma _\mu \ n_\mu ,$ y  $n=(n_\mu)$ es el vector
unitario normal exterior a $\partial M$.

En particular, si $f=T\varphi $ con $\varphi \in {\rm Ker} (D)$, se tiene que
$QT\varphi=T\varphi$, como corresponde a la definici\'on de $Q$.

Si bien $Q$ no es \'unico, ya que se puede tomar cualquier soluci\'on
fundamental de $D$, su imagen y su s\'{\i}mbolo principal, $q(x;\xi )$, son
independientes de la elecci\'on de $K( x,y)$\cite{Calderon}.

Tomando expl\'{\i}citamente el l\'{\i}mite en (\ref{poto}) tenemos
\begin{equation}
\label{104}
\begin{array}{c}
Q f(x)=\frac 12f(x)-i\ P.V.\int_{\partial M}K_0(x,y)
\not \! n\ f(y)\ d\sigma _y\  \\  \\
-i\ \int_{\partial M}R(x,y)\ \not \! n\ f(y)\ d\sigma _y  .
\end{array}
\end{equation}

Debido a su comportamiento local, $R(x,y)$ da lugar a un pseudo-diferencial de
orden $\leq -1$. Entonces, s\'olo los dos primeros t\'erminos del miembro
derecho de (\ref{104}) contribuyen al s\'{\i}mbolo principal de $Q$, que
resulta ser\cite{fat}
\begin{equation}
\label{q}
q(x;\xi )=\frac 12(Id_{k\times k}+i\ \frac{\not \! \xi }{\vert \xi \vert } \not
\! n).
\end{equation}

N\'otese que $q(x;\xi )$ es una matriz de $k \times k$ que satisface
\begin{equation}
\label{rango}
\begin{array}{c}
q(x;\xi )\ q(x;\xi )=q(x;\xi ) \\  \\
tr\ q(x;\xi )={k/2},
\end{array}
\end{equation}
de modo que su rango es $rank\, q(x;\xi )=k/2.$

\bigskip

En particular, si $\nu=2$ y las matrices de Dirac est\'an dadas por
\begin{equation}
\label{gm}
\begin{array}{c}
\gamma _0=\sigma _1=\left(
\begin{array}{cc}
0 & 1 \\
1 & 0
\end{array}
\right) ,\quad \gamma _1=\sigma _2=\left(
\begin{array}{cc}
0 & -i \\
i & 0
\end{array}
\right) , \\
\\
\gamma _5=-i\gamma _0\gamma _1=\sigma _3=\left(
\begin{array}{cc}
1 & 0 \\
0 & -1
\end{array}
\right) ,
\end{array}
\end{equation}
obtenemos
\begin{equation}
\label{q2}q(x;\xi )=\left(
\begin{array}{cc}
H(\xi ) & 0 \\
0 & H(-\xi )
\end{array}
\right) , \quad \forall x\in \partial M,
\label{Heavi}
\end{equation}
 donde $H(\xi )$ es la funci\'on de Heaviside.

\bigskip

Supondremos que $rank\ q(x;\xi )=r$ es constante (\'ese es siempre el caso para
$\nu \geq 3$, y tambi\'en vale en el ejemplo anterior con $\nu=2$).

De acuerdo con Calder\'on\cite{Calderon}, las {\it condiciones de contorno
el\'{\i}pticas} pueden ser caracterizadas en t\'erminos de $q(x;\xi )$, seg\'un
la siguiente definici\'on.

Sea $B$ un operador pseudo-diferencial de orden cero,
\be
B:[L^2(\partial M,E_{/\partial
M})]\rightarrow [L^2(\partial M,G)],
\ee
donde $G$ es un fibrado complejo de dimensi\'on $r$.
$B$ da lugar a una condici\'on de contorno el\'{\i}ptica para un operador
diferencial de primer orden $D$ si,
\begin{equation}
\label{erre}
\forall \xi :\vert \xi \vert  \geq 1, \quad
rank(b(x;\xi )\, q(x;\xi ))=rank(q(x;\xi ))=r ,
\end{equation}
donde $b(x;\xi )$ coincide con el s\'{\i}mbolo principal de $B$ para $\vert \xi
\vert \geq 1.$

En ese caso se dice que
\begin{equation}
\label{BoundaryProblem}\left\{
\begin{array}{c}
D\varphi =\chi\ \
\rm{ en }\ M \\  \\
BT\varphi =f\ \rm{ sobre }\ \partial M
\end{array}
\right.
\end{equation}
es un {\it problema el\'{\i}ptico de borde}.

En particular, la condici\'on (\ref{erre}) implica que, para cada $s\in {\bf
R},$ la imagen de $B Q$ es un subespacio cerrado del espacio de
Sobolev\footnote{El espacio de Sobolev $H^s({\rm \bf R}^{\nu})$ es el
completamiento del espacio de Schwartz ${\cal S}({\rm \bf R}^{\nu})$ respecto
de la norma $|| u(x) ||_s = || (1+|\xi|^2)^{s/2}\, \hat{u}(\xi) ||_{{\rm \bf
L}_2}$. \hfill\break \hfill} $H^s(\partial M,G)$, de codimensi\'on
finita\cite{Calderon}.

Un problema el\'{\i}ptico como (\ref{BoundaryProblem}) tiene una soluci\'on
$\varphi \in H^1(M,E)$ para cada $(\chi ,f)$ en un subespacio de
$L^2(M,E)\times H^{1/2}(\partial M,G)$ de codimensi\'on finita. Adem\'as, esa
soluci\'on es \'unica a menos de una funci\'on en un subespacio nulo de
dimensi\'on finita\cite{Calderon}. En otras palabras, el operador
\begin{equation}
(D,BT):H^1(M,E)\rightarrow L^2(M,E)\times H^{1/2}(\partial M,G)
\end{equation}
es Fredholm.

\bigskip

Se denota por $D_B$ a la clausura del operador  $D$ actuando sobre funciones
$\varphi $ $\in C^\infty (M,E)$ que satisfacen la condici\'on de contorno
homog\'enea $B(T\varphi )=0.$

\subsection*{Condiciones de contorno locales}

Cuando $B$ es un operador {\it local}, la anterior definici\'on conduce a las
condiciones el\'{\i}pticas locales cl\'asicas, tambi\'en llamadas condiciones
de Lopatinsky - Shapiro.

Para el operador de Dirac antes considerado, se tienen condiciones locales
cuando la acci\'on de $B$ corresponde a la multiplicaci\'on por una matriz
de $k/2 \times k$,
cuyos elementos son funciones definidas sobre $\partial M$.

\bigskip

Pero no siempre es posible imponer condiciones el\'{\i}pticas locales. Por
ejemplo, el operador de Dirac {\it quiral} en dimensi\'on par no las admite.

En dimensi\'on $\nu=4$, y tomando las matrices de Dirac cerca del borde como
\begin{equation}
\gamma _n=i\left(
\begin{array}{cc}
0 & Id_{2\times 2} \\
-Id_{2\times 2} & 0
\end{array}
\right) \ \ \ \rm{and}\quad \gamma _j=\left(
\begin{array}{cc}
0 & \sigma _j \\
\sigma _j & 0
\end{array}
\right) ,
\end{equation}
donde $ j=1,2,3$ corresponde a las direcciones tangentes al borde,
el s\'{\i}mbolo principal del operador de Calder\'on correspondiente al
operador de Dirac completo tiene la forma
\begin{equation}
q(x;\xi )=\frac 12\left(
\begin{array}{cc}
Id_{2\times 2}+\dfrac{\xi .\sigma }{\vert \xi \vert } & 0 \\
0 & Id_{2\times 2}-\dfrac{\xi .\sigma }{\vert \xi \vert }
\end{array}
\right) .
\label{q4}
\end{equation}

El s\'{\i}mbolo principal del operador de Calder\'on para el operador de Dirac
quiral corresponde al bloque superior izquierdo de (\ref{q4}),
\begin{equation}
q_{ch}(x;\xi )=\frac 12\left(
\begin{array}{cc}
1+\dfrac{\xi _3}{\vert \xi \vert } & \dfrac{\xi _1-i\xi _2}{\vert \xi \vert }
\\
  &  \\
\dfrac{\xi _1+i\xi _2}{\vert \xi \vert } & 1-\dfrac{\xi _3}{\vert \xi \vert }
\end{array}
\right) .
\end{equation}
As\'{\i}, $q_{ch}(x;\xi )$ es una matriz de 2$\times 2$ autoadjunta,
idempotente y de traza 1. En consecuencia su rango es
$rank \, q_{ch}(x;\xi )=1$.

Ahora bien, si existiera una condici\'on de contorno local para este problema,
con un  $B$ cuyo s\'{\i}mbolo principal est\'e dado por $b(x)=$
$(\beta _1(x),$$\beta _2(x))$, deber\'{\i}a ser $rank(b(x)\, q_{ch}(x;\xi
))=1,\ \forall \xi \neq 0.$ Sin embargo, es f\'acil ver que para
\begin{equation}
\xi _1=\frac{-2\beta _1\beta _2}{\beta _1^2+\beta _2^2},\qquad \xi _2=0\quad
\rm{and\quad }\xi _3=\frac{\beta _2^2-\beta _1^2}{\beta _1^2+\beta _2^2},
\end{equation}
se tiene que $rank(b(x)\, q_{ch}(x;\xi ))=0.$

Esto es un ejemplo de lo que se conoce como {\it obstrucciones topol\'ogicas} a
la existencia de condiciones locales el\'{\i}pticas.
(No obstante, siempre existen condiciones de contorno  locales
e\-l\'{\i}p\-ti\-cas
para el operador de Dirac {\it completo}.)

\subsection*{Condiciones de contorno globales}

Una clase de condiciones de contorno el\'{\i}pticas {\it no locales} para
operadores de Dirac corresponde a las introducidas por M.\ Atiyah, V.\ Patodi y
I. Singer\cite{APS} en el marco del {\it teorema del \'{\i}ndice para
variedades con borde}.

Cerca del borde, el operador de Dirac quiral tiene la forma
\begin{equation}
\varrho \ (\partial _t+{\cal A}),
\end{equation}
donde $t$ es la coordenada normal interior a $\partial M$,
\be
{\cal A}:L^2(\partial M,E_{/\partial M})\rightarrow L^2(\partial M,E_{/\partial
M})
\ee
es un operador autoadjunto, y $\varrho:E\rightarrow F$ es un isomorfismo
isom\'etrico.

El operador $P_{APS}$, que define las {\it condiciones de contorno globales},
es el proyector ortogonal sobre el subespacio cerrado de $L^2(\partial
M,E_{/\partial M})$ que subtienden las autofunciones de ${\cal A}$
correspondientes a  autovalores no nagativos:
\be
P_{APS}=\sum_{\lambda \geq 0} \phi_\lambda (\phi_\lambda , \cdot ),\quad{\rm
donde\  }{\cal A}\phi_\lambda = \lambda \phi_\lambda.
\ee
$P_{APS}$ es un pseudo-diferencial de orden cero, cuyo s\'{\i}mbolo principal
{\it coin\-cide} con el del correspondiente operador de
Calder\'on\cite{Booss-W}.

De ese modo, su composici\'on con $q_{ch}(x,\xi)$ tiene el mismo rango que
$q_{ch}(x,\xi)$, y define una condici\'on de contorno el\'{\i}ptica.

El problema
\begin{equation}
\label{BPro}\left\{
\begin{array}{c}
D\varphi =\chi\ \
\rm{ en }\ M \\  \\
P_{APS}T\varphi =f\ \rm{ sobre }\ \partial M
\end{array}
\right.
\end{equation}
tiene una soluci\'on $\varphi \in H^1(M,E)$  para cada $(\chi ,f)$, con $\chi $
en un subespacio de $L^2(M,E)$ de codimensi\'on finita, y $f$  en la
intersecci\'on de $H^{1/2}(\partial M,E_{/\partial M})$ con la imagen de
$P_{APS}$. Esa soluci\'on es \'unica a menos de funciones en un subespacio nulo
de dimensi\'on finita.

Como la codimensi\'on de $P_{APS}\ [L^2(\partial M,E_{/\partial M})]$ no es
finita, el operador
\begin{equation}
(D,P_{APS}T):H^1(M,E)\rightarrow L^2(M,E)\times H^{1/2}(\partial
M,E_{/\partial M})
\end{equation}
no es Fredholm.

\bigskip

N\'otese que, si bien  $P_{APS}$ y $Q$ tienen el mismo s\'{\i}mbolo principal,
puede decirse que sus acciones {\it van en sentidos opuestos}.
El operador de Calder\'on $Q$ est\'a relacionado con la existencia de
soluciones del operador diferencial en el interior de la variedad $M$.
Por su parte, el proyector $P_{APS}$ es tal que limita el Ker$(D_{P_{APS}})$ al
conjunto de soluciones de $D$ que admiten una extensi\'on de cuadrado
integrable sobre la elongaci\'on no compacta de $M$, obtenida pegando a su
borde el cilindro semi-infinito $(-\infty,0] \times \partial M$.

\bigskip

Por ejemplo, para el operador de Dirac completo en un espacio Eucl\'{\i}deo
bidimensional, vemos que el problema de borde (autoadjunto)
\begin{equation}
\begin{array}{c}
\left(
\begin{array}{cc}
0 & D^{\dagger } \\
D & 0
\end{array}
\right) \left(
\QDATOP{\varphi_1}{\varphi_2}\right) =\left(
\QDATOP{\chi_1}{\chi_2}\right)
\ \  \rm{ en }\ M, \\  \\
(P_{APS},\varrho (I-P_{APS})\ \varrho ^{*})\left(
\QDATOP{\varphi_1}{\varphi_2}\right) =
\left( \QDATOP{f_1}{f_2}\right)\ \ \rm{
sobre }\ \partial M,
\end{array}
\end{equation}
es el\'{\i}ptico.

En efecto, como se se\~nal\'o antes,  $P_{APS}$ y $Q$ tiene el mismo
s\'{\i}mbolo principal. Para el bloque quiral
\be
D=i\left( \partial_0 + i \partial_1\right) + \left(A_0 + i A_1 \right),
\ee
de (\ref{Heavi}) se tiene
\be
\sigma _0 (P_{APS})(x,\xi)=H(\xi ),
\ee
y para su adjunto
\be
\sigma _0(\varrho\, (I-P_{APS})\, \varrho ^{*})=H(-\xi ).
\ee
Entonces, el s\'{\i}mbolo principal de $B=(P_{APS},\varrho\, (I-P_{APS})\,
\varrho ^{*})$ es $b(x;\xi )=(H(\xi ),H(-\xi ))$, y satisface
\begin{equation}
rank(b(x;\xi )\ q(x;\xi ))=rank(q(x;\xi ))\qquad \forall \xi \neq 0.
\end{equation}

\subsection*{Determinantes funcionales de operadores el\'{\i}pticos en
variedades con borde}

La definici\'on de Seeley para las potencias complejas de operadores
el\'{\i}pticos se generaliza, bajo ciertas condiciones, al  caso de problemas
el\'{\i}pticos de borde con condiciones {\it locales}\cite{Seeleycb}.

En ese caso, para $\Re(s)>\nu$ (ya que $\omega=1$), la potencia $D_B^{-s}$ es
un operador integral cuyo n\'ucleo, $J_{-s}(x,y)$, es continuo en $(x,y,s)$, y
anal\'{\i}tico en $s$.

Para $\Re(s)$ suficientemente grande, se define la funci\'on
\be
\zeta _{(D_B)}(s)= Tr\{D_B^{-s}\}= \int_M d^\nu x \,
tr\left\{J_{-s}(x,x)\right\}.
\ee
$\zeta _{(D_B)}(s)$ se extiende anal\'{\i}ticamente a una funci\'on meromorfa,
cuyas \'unicas singularidades son polos simples situados en $s=\nu-j$,
$j=0,1,2, ...$, con residuos nulos en $s=0,-1,-2, ...$

En particular, la funci\'on $\zeta _{(D_B)}(s)$ es regular en $s=0$, por lo que
puede definirse un determinante regularizado para $D_B$ como\footnote{ Este
determinante (invariante de gauge) tambi\'en puede ser expresado en t\'erminos
de la funci\'on de Green del problema de borde como en \cite{a1,a2}}
\begin{equation}
\label{DR}Det\ (D_B)=\exp [-\frac d{ds}\ \zeta _{(D_B)}(s)]\vert _{s=0} .
\end{equation}

Pero la construcci\'on de las potencias complejas de problemas
e\-l\'{\i}p\-ti\-cos de borde con condiciones de contorno {\it globales} es un
problema a\'un abierto\cite{SG}, por lo que en ese caso no se puede recurrir a
una definici\'on similar.

\section*{El operador de Dirac en $\nu=2$ con condiciones de contorno globales}

Presentamos ahora el c\'alculo del deter\-mi\-nante funcional
para un modelo simple de un operador de Dirac en un disco, bajo
condiciones de contorno globales {\it (o espectrales)},
 y en presencia de un flujo
Abeliano con simetr\'\i a cil\'\i ndrica\cite{global}. En\cite{Sitenko,Moreno}
y las referencias all\'\i \  citadas
puede encontrarse trabajo relacionado con el problema aqu\'\i \ tratado.

Consideramos el
operador  $D=\ \ \not \!\!\!\!i\partial
+\not \!\!\!A$, definido sobre el espacio de
funciones di\-fe\-ren\-cia\-bles en un disco, que satisfacen condiciones de
contorno  globales en $r=R$.

El  campo $A_\mu $  en el gauge de Lorentz puede escribirse
como $A_\mu =\epsilon _{\mu \nu }\ \partial_\nu \phi$.
Eligiendo para $\phi$ una funci\'on suave que s\'olo dependa de $r$,
$\phi (r)$, el campo de gauge resulta
\be
A_r=0, \qquad A_\theta(r)=-\partial_r\phi
(r)=-\phi^{^{\prime }}(r).
\ee

 En esas condiciones, el {\it flujo} a
trav\'es del disco est\'a dado por
\begin{equation}
\label{3.2}
{\k } =\dfrac{\Phi}{2\pi}=\dfrac{1}{2\pi}\oint_{r=R}\ A_\theta
\ R\ d\theta =- R\phi^{^{\prime }}(R).
\end{equation}

Con las convenciones usuales para las matrices $\gamma_\mu$, el
operador de Dirac puede escribirse como
\begin{equation}
 D=e^{-\gamma _5\phi
(r)\ }i\not \! \partial \ e^{-\gamma _5\phi (r)}
=\left(
\begin{array}{cc} 0 & \varrho
^{-1}(\partial _r+{\cal A}) \\ \varrho \ (-\partial _r+{\cal A}) & 0
\end{array} \right),
\label{op}
\end{equation}
donde
\begin{equation}
\varrho =-\ i\ e^{i\theta },
\end{equation}
y
\begin{equation}
{\cal A}(r)=-\frac ir\ \partial _\theta +\ \partial
_r\phi (r).
\label{beder}
\end{equation}
Sobre el borde,  el problema de autovalores para el
operador autoadjunto ${\cal A}(R)$ es
\be
{\cal A}(R)\,e^{in\theta } =a_n\, e^{in\theta }, \quad {\rm con\ }a_n = \frac
1R(n-{\k}), \   n \in {\rm \bf Z}.
\ee

Las condiciones de contorno de Atiyah, Patodi y Singer est\'an
caracterizadas por las proyecciones
\begin{equation}
\label{apsbc}\left(
\begin{array}{cc}
{\cal P}_{\geq} & \varrho\, (1-{\cal P}_{\geq})\, \varrho ^{*}
\end{array}
\right) \left( \QATOP{\varphi (R,\theta )}{\chi (R,\theta )}\right) =0,
\end{equation}
donde ${\cal P}_{\geq}$ proyecta sobre el subespacio de autofunciones de ${\cal
A}(R)$ con autovalores $a_n = \frac 1R(n-{\k}) \geq 0$.

\bigskip

Si $k$ es un entero tal que $k<{\k } \leq k+1$,
\begin{equation}
{\cal P}_{\geq}=\frac 1{2\pi }\sum_{n\geq k+1}e^{in\theta }\left(
e^{in\theta
},\ \cdot \ \right),
\end{equation}
\begin{equation}
\varrho\, (1-{\cal P}_{\geq})\, \varrho^{*}=\varrho \, {\cal P}_{<}\,
\varrho^{*}
=\frac
1{2\pi }\sum_{n\leq k+1}e^{in\theta }\left( e^{in\theta
},\ \cdot \ \right) ={\cal P}_{\leq}.
\end{equation}

El operador definido por (\ref{op}) y (\ref{apsbc}), que denotaremos por
$\left( D\right) _{{\ka }}$, resulta autoadjunto.

Nuestra intenci\'on es calcular el cociente de los
determinantes de los operadores  $\left( D\right) _{{\ka} }$ y
$(\ \not \!\!\!\!i\,\partial )_{{\ka }=0}$. Dado que las condiciones
de contorno globales (\ref{apsbc}) dependen del flujo de manera
discontinua, vamos a proceder en dos pasos:
\begin{equation}
\left( D\right) _{{\ka } }\rightarrow (i\not \! \partial
)_{{\ka }}\rightarrow (i\not \! \partial  )_{{\ka }=0}  .
\end{equation}
En el primer paso, manteniendo fijas las condiciones de
contorno, multiplicaremos $A_\mu$ por un par\'ametro $\alpha$
que variaremos continuamente entre 1 y 0:
\begin{equation}
\label{opalf}
D_\alpha =i\not \! \partial +\alpha \not \! \!
A=e^{-\alpha \gamma
_5\phi (r)\;\ }i\not \! \partial \ e^{-\alpha \gamma _5\phi (r)}.
\end{equation}
Aqu\'{\i} ser\'a necesario el conocimiento de la funci\'on de
Green del problema. El segundo paso ser\'a posible a partir del
c\'alculo expl\'{\i}cito de los autovalores de $(i\not \!
\partial  )_{\ka }$.

Pero hay una dificultad adicional, ya que las condiciones de
contorno (\ref{apsbc}) admiten la existencia de $\vert k+1
\vert $ autovectores linealmente independientes de $(D_\alpha
)_{{\ka }}$ co\-rres\-pondientes a autovalores nulos. Para $ k\geq
0$ dichos autovectores est\'an dados por\cite{global}
\begin{equation}
\dfrac{\ e^{\alpha \phi (r)}}{\sqrt{2\pi \ q_n(R;\alpha)}}\ \left(
\QDATOP{X^n}{0}
\right) ,{\rm \ \   para }\ 0\leq n\leq k,
\label{ceros}
\end{equation}
donde $X=x_0 + i x_1=r e^{i \theta}$, y el factor de
normalizaci\'on es
\begin{equation}
q_n(u;\alpha)=\dint_0^u\ e^{2\alpha \phi (r)}\ r^{2n+1}\ dr .
\end{equation}
Para $k<-1$ se tienen expresiones similares, pero con quiralidad opuesta.
N\'otese que para $k=-1$ (en particular, para $\Phi = 0$) no
hay modos cero. En lo que sigue, por simplicidad, s\'olo nos
ocuparemos del caso $k\geq -1$.

El n\'ucleo del proyector ortogonal sobre Ker$(D_\alpha )_{{\ka
}}$, $P_\alpha $, est\'a dado por
\begin{equation}
P_\alpha (z,w)=\sum\limits_{n=0}^k\dfrac{\ e^{\alpha
[\phi (z)+\phi (w)]}}{
2\pi \ q_n(R;\alpha)}\left(
\begin{array}{cc}
(ZW^{*})^n & 0 \\
0 & 0
\end{array}
\right) .
\end{equation}
Ahora bien, siendo $(D_\alpha +P_\alpha )_{{\ka }}$
invertible, definimos
\begin{equation}
Det^{^{\prime }}(D_\alpha )_{{\ka }}\equiv Det(D_\alpha +P_\alpha
)_{{\ka }},
\end{equation}
y escribimos
\begin{equation}
\label{dobleco}
\frac{Det^{^{\prime }}(D)_{{\ka }}}{Det(i\not \! \partial
)_{{\ka }=0}}=\frac{Det(D+P_1)_{{\ka }}}{Det(i\not \! \partial
+P_0)_{{\ka }}}\ \frac{Det(i\not \! \partial  +P_0)_{{\ka }}}{Det(i\not \!
\partial  )_{{\ka }=0}}.
\end{equation}

Para calcular el primer cociente del segundo miembro de
(\ref{dobleco}) tomamos la derivada
\begin{equation}
\label{dalfa}
\dfrac \partial {\partial \alpha }\left[\ln Det{(D_{\alpha}
+P_{\alpha})_{{\ka
}}}\right]=Tr \left[ (\not \! \! A+ \partial_{\alpha}P_{\alpha} )
G_\alpha
\right],
\end{equation}
donde $G_{\alpha}(x,y)$ es la funci\'on de Green del problema
\begin{eqnarray}
(D_\alpha +P_\alpha )\ G_{\alpha}(x,y)=\delta (x,y), \nn
\\
\left(
\begin{array}{cc}
{\cal P}_{\geq} & \, \,  {\cal P}_{\leq}
\end{array}
\right) G_{\alpha}(x,y)\vert _{r=R }=0.
\end{eqnarray}

Dado que $(D_{\alpha})_{{\ka}}$ es autoadjunto,
$G_{\alpha}(x,y)$ tiene la estructura
\begin{equation}
\label{green2}
G(x,y)_{\alpha}=(1-P_\alpha )\ {\cal G}_{\alpha}(x,y)\
(1-P_\alpha )+\ P_\alpha,
\end{equation}
donde ${\cal G}_{\alpha}(x,y)$ es el n\'ucleo de la inversa a
derecha de  $(D_\alpha)_{{\ka}} $ en el complemento ortogonal
de $Ker(D_\alpha )_{{\ka}}$\cite{global},
\begin{equation}
{\cal G}_{\alpha}(x,y)=\frac 1{2\pi i} \times
\left(
\begin{array}{cc}
0 & \frac{e^{\alpha [\phi (x)-\phi (y)]}}{X-Y}\left(
\frac XY\right) ^{k+1}
\\ \frac{e^{-\alpha [\phi (x)-\phi (y)]}}{X^{*}-Y^{*}}\left(
\frac{Y^{*}}{%
X^{*}}\right) ^{k+1} & 0
\end{array}
\right) ,
\end{equation}
expresi\'on que, remplazada en (\ref{green2}), permite obtener
 $G(x,y)_{\alpha}$.

Como $P_\alpha$ es un  proyector ortogonal,
\be
(P_\alpha)^2=P_\alpha, \qquad
\frac{\partial P_\alpha}{\partial \alpha} \left( 1 - P_\alpha\right) =
P_\alpha \frac{\partial P_\alpha}{\partial \alpha},
\ee
de (\ref{green2})
resulta que $Tr \left[ (\partial_\alpha P_\alpha )\, G_\alpha
\right]=0$. De ese modo, (\ref{dalfa}) se reduce al c\'alculo de
$Tr \left[ \not  \! \! A G_\alpha \right]$.

El argumento de esa
traza tiene un n\'ucleo singular en la diagonal principal; para
darle un sentido recurriremos a un {\it point-splitting},
definiendo
\[
Tr \left[ \not  \! \! A G_\alpha \right]=
\]
\begin{equation}
{\rm lim.\ sim.\ }_{\epsilon \rightarrow 0}\int_{r<R} d^2x \, \,
tr\left[ \not
\! \! A(x) G_\alpha(x,x+\epsilon) e^{i \alpha \epsilon \cdot A(x)}
\right] ,
\end{equation}
donde hemos agregado el factor de Schwinger para ga\-ran\-ti\-zar la
invarianza de gauge del resultado, y entendemos por l\'{\i}mite
sim\'etrico  la semisuma de los l\'{\i}mites laterales.

Integrando en $\alpha$ entre 0 y 1 obtenemos\cite{global}
\[
\ln
\left[
\frac{Det(D+P_1)_{{\ka }}}{Det(i\not \! \partial +P_0)_{{\ka }}}
\right] =-\frac
1{2\pi }\int _{r<R} d^2x\  {\phi ^{\prime }}^2
\]
\begin{equation}\label{a1}
 -2\ (k+1)\ \phi (R)\
+\sum\limits_{n=0}^k\
\ln \left[2(n+1)\frac{q_n(R;1)}{R^{2(n+1)}}\right]\ .
\end{equation}
\hfill\break
N\'otese que cuando no hay modos cero ($k+1=0$) el resultado se
reduce s\'olo al primer t\'ermino.

\bigskip

En lo que sigue evaluaremos el segundo cociente en
el segundo miembro de
(\ref{dobleco}) en t\'erminos de la funci\'on $\zeta$ del
operador  ${(\ \not \!\!\!i\partial  +P_0)_{{\ka }}}$. Para
ello calcularemos ex\-pl\'{\i}\-ci\-ta\-men\-te su espectro.

Las autofunciones de ${(\ \not \!\!\!i\partial  +P_0)_{{\ka }}}$ son de la
forma
\begin{equation} \psi
_n(r,\theta)=\left( \QATOP{\varphi_n (r,\theta )}{\chi_n (r,\theta )}\right)
=\left( \QATOP{J_n(\vert \lambda \vert r)\
e^{in\theta }}{-i\frac{\vert \lambda \vert} { \lambda
}J_{n+1}(\vert \lambda \vert r)\ e^{i(n+1)\theta }}\right),
\label{autofun}
\end{equation}
y satisfacen las condiciones de contorno
\begin{equation}
{\cal P}_{\geq}\, \varphi_n (R,\theta ) =
 \frac 1{2\pi }\sum_{n\geq k+1}e^{in\theta }\left(
e^{in\theta
}, \varphi_n (R,\theta )\right) =0,
\end{equation}
\begin{equation}
{\cal P}_{\leq} \,\chi_n (R,\theta )
=\frac 1{2\pi }\sum_{n\leq k+1}e^{in\theta }\left( e^{in\theta
},\chi_n (R,\theta )\right) =0 .
\end{equation}
Para $n\geq
k+1$ el
autovalor es  $\lambda =\pm j_{n,l}/R$  ( donde $j_{n,l}$ es el
$l$-th cero de $J_n(z)$). An\'alogamente, para $n \leq k$,
$\lambda =\pm j_{n+1,l}/R.$

En consecuencia, los
autovalores son
\begin{equation} \lambda _{n,l}=\pm j_{n,l}/R,\
{\rm para\ }n=0,\pm 1,\pm 2,...\ {\rm y\ }\ l=1,2,...\ ,
\end{equation}
donde $j_{-n,l}=j_{n,l} $. N\'otese que para
$n=k+1$ los autovalores tienen multiplicidad dos, mientras que
para $n\neq k+1$ su multiplicidad es s\'olo uno.

Para $\Re (s)$ suficientemente grande, se puede
definir\cite{global}
\[
\displaystyle \zeta _{(i\not\partial  +P_0)_{{\ka }}}(s)=\vert k+1\vert
+(1+e^{-i\pi s}) \times
\]
\begin{equation}
\label{zeta}
\left\{ \sum\limits_{n=-\infty }^{\infty }\sum\limits_{l=1}^\infty \
\left(
\dfrac{
j_{n,l}}R\right) ^{-s}+\sum\limits_{l=1}^\infty \ \left( \dfrac{j_{\vert
k+1\vert,l}}%
R\right) ^{-s}\right\} .
\end{equation}
\hfill\break
El primer t\'ermino, $\vert k+1\vert$, es la dimensi\'on del
sub\-es\-pa\-cio Ker${(\ \not \!\!\!i\partial )_{{\ka }}}.$
La doble suma en el segundo t\'ermino (que es independiente de $k$) coincide
con la
funci\'on $\zeta$ del Laplaciano en el disco sometido a las
condiciones de contorno de Dirichlet (locales!). Ello garantiza que su
extensi\'on anal\'{\i}tica sea regular en $s=0$.
La extensi\'on anal\'{\i}tica de la segunda suma,
\begin{equation}
f_\nu (s)\equiv \sum_{l=1}^\infty \ (\ j_{\nu,l})^{-s}.
\end{equation}
tambi\'en resulta regular en el origen, por lo que
$\zeta _{(i\not\partial  +P_0)_{{\ka }}}(s)$ es anal\'{\i}tica
en $s=0$.

En el marco de la regularizaci\'on {\it \`a la } $\zeta$, el cociente de
de\-ter\-mi\-nan\-tes queda definido por
\[
\ln \left[
\dfrac{Det(i\not \! \partial  +P_0)_{\ka }}{Det(i\not \!
\partial  )_{{\ka }
=0}}
\right] \equiv
-\dfrac d{ds}\left[ \zeta _{(i\not\partial  +P_0)_{{\ka }}}(s)-\zeta
_{(i\not\partial  )_{{\ka=0 }}}(s)\right] _{s=0} =
\]
\begin{equation}
\label{qdl}
-2\left[ f_{\vert k+1\vert}^{\prime }(0)-f_0^{\prime }(0)+
(\ln R-\frac{i\pi }
2)[f_{\vert k+1\vert}(0)-f_0(0)]\right] .
\end{equation}
Recurriendo al desarrollo asint\'otico de los ceros de las
funciones de Bessel se encuentra
\begin{equation}
\label{f0}f_\nu (0)=-\frac \nu 2-\frac 14,
\end{equation}
y
\begin{equation}
\label{f'0}
f_\nu ^{\prime }(0)=-\frac 12\ln 2+\left( \frac{2\nu -1}4\right)
(\ln \pi -\gamma )
-\sum\limits_{l=1}^\infty \ln \left[ \frac{\ j_{\nu ,l}}{%
l\pi }\ e^{-\left( \frac{2\nu -1}{4\ l}\right) }\right] ,
\end{equation}
donde $\gamma $ es la constante de Euler.

Finalmente, teniendo en cuenta que cada parte ha sido
regularizada de manera invariante de gauge, obtenemos\cite{global}
\[
\displaystyle \ln\left[
\frac{Det(D+P_1)_{{\ka }}}{Det(i\not \! \partial )_{{\ka }=0}}
\right] =
-\frac
1{2\pi }\int_{r<R} d^2x\ A_\mu (\delta_{\mu \nu}-
\dfrac{\partial_{\mu}\partial_{\nu}}{\partial^2}) \ A_\nu
\]
\[
 -2\ (k+1)\ \phi(R) \displaystyle +\sum\limits_{n=0}^k\
\ln\left[2(n+1)\frac{q_n(R;1)}{R^{2(n+1)}} \right]
\]
\begin{equation}
\label{final}
 -\vert k+1\vert [\frac{i\pi }2-\gamma -\ln (\frac R\pi)]
+2\sum\limits_{l=1}^\infty \ln \left[ \frac{\ j_{\vert k+1\vert,l}}{\
j_{0,l}}\
e^{-\left( \frac{\vert k+1\vert}{2\ l}\right) }\right] .
\end{equation}

\subsection*{Relaci\'on con el teorema del \'\i ndice}

La variaci\'on del determinante funcional frente a una
transformaci\'on axial  global,
\[
 e^{-\gamma_{5} \epsilon }
(D+P_1)_{{\ka }}e^{-\gamma_{5} \epsilon }
=(D+e^{-\gamma_{5} \epsilon }P_1e^{-\gamma_{5} \epsilon })_{{\ka }} ,
\]
\begin{equation}
e^{-\gamma_{5} \epsilon }  (i\not \! \partial )_{{\ka }=0} \, e^{-\gamma_{5}
\epsilon }=(i\not \! \partial )_{{\ka }=0}
\end{equation}
(con $\epsilon$ constante), est\'a relacionada con el {\it
\'{\i}ndice} del o\-pe\-ra\-dor de Dirac:
\begin{equation}
\label{dep}
\displaystyle \dfrac{\partial}{\partial \epsilon}\ln\left[
\frac{Det\left(e^{-\gamma_{5} \epsilon }(D+P_1)_{{\ka }}e^{-\gamma_{5}
\epsilon
}\right)}
{Det\left( e^{-\gamma_{5} \epsilon }(i\not \! \partial )_{{\ka
}=0}e^{-\gamma_{5} \epsilon }\right) }
\right]  =
 -2Tr \left[ \gamma _{5} P_1  \right]=-2(N_{+}-N_{-}),
\end{equation}
donde $N_{+(-)}$ es el n\'umero de modos cero con quiralidad
positiva (negativa).

Puede verse que nuestra estrategia de c\'alculo conduce al
valor correcto para esa diferencia.

En efecto, teniendo en
cuenta que la inversa del operador transformado est\'a dada por
\begin{equation}\label{green3}
G^{(\epsilon)}_{\alpha}(x,y)=(1-P_{\alpha} )\ {\cal G}_{\alpha}(x,y)\
(1-P_{\alpha} )+e^{\gamma_{5} \epsilon }\
P_{\alpha}\,  e^{\gamma_{5} \epsilon }
\end{equation}
en lugar de (\ref{green2}), la \'unica diferencia se presenta
en el primer t\'ermino del miembro derecho de (\ref{zeta}) (que cuenta la
dimensi\'on de Ker$(i\not \! \partial )_{\ka } $), que ahora
aparece multiplicado por un factor $e^{\pm 2\epsilon s}$. Tomando
la derivada respecto de $\epsilon$ (en $s=0$) obtenemos el valor correcto para
el \'{\i}ndice,
\begin{equation}
\label{enemas}
{\rm index }(D)_{{\ka }}= N_{+}  -N_{-}= k+1.
\end{equation}

\bigskip

El teorema de Atiyah, Patodi y Singer relaciona el valor de
${\rm index }(D)_{{\ka }}$ con la {\it  asimetr\'{\i}a
espectral} del o\-pe\-ra\-dor de Dirac restringido al borde,
\be
{\cal A}={\cal A}(R)=-\frac i R \ \partial _\theta +\ \partial _r \phi (R)
\ee
A partir de sus autovalores, $a_n = \frac 1R(n-{\k})$, se define
\begin{equation}
\eta_{({\cal A})}(s)= R^s \sum_{n \neq {\k }} {
sig(n-{\k } ) \, \,  \vert n-{\k } \vert^{-s}},
\end{equation}
serie convergente para $\Re (s) > 1$.

Su extensi\'on
anal\'{\i}tica a $s=0$ est\'a dada por\cite{global}
\begin{equation}
\eta_{({\cal A})}(0)=2({\k }-k)-1-{ h}({\cal A}),
\end{equation}
donde $h({\cal A})=$ dim Ker$({\cal A})$.

Siguiendo la construcci\'on de\cite{APS}, y teniendo
en cuenta que en nuestro caso $\varrho =\varrho(\theta) =-\ i\
e^{i\theta }$, resulta
\begin{equation}
{\rm index}\,  D= {\k } + \frac{ \left[ 1- { h}({\cal A})-
\eta_{({\cal A})}(0)\right]}{2}=k+1,
\end{equation}
en coincidencia con  (\ref{enemas}). El primer t\'ermino de la
expresi\'on intermedia corresponde a la bien conocida
contribuci\'on de volumen\cite{Annals}
. El segundo es la
contribuci\'on del borde hallada en\cite{APS}, pero corrida en
$1/2$. Esta diferencia, producto de la dependencia con $\theta$
del  factor $\varrho$, ya hab\'{\i}a sido
encontrada en\cite{Sitenko} para condiciones de contorno
espectrales ligeramente diferentes de las aqu\'{\i} empleadas.

\subsection*{Conclusiones}

Hemos podido evaluar el determinante funcional de un operador de Dirac en un
disco, en presencia de un flujo (Abeliano) con simetr\'{\i}a axial, y sometido
a las condiciones de contorno globales introducidas por Atiyah, Patodi y
Singer.

En un primer paso, manteniendo fijas las condiciones de contorno, hemos
variado el campo de gauge. En esta parte del c\'alculo, que reposa en el
conocimiento de la funci\'on de Green del problema, hemos empleado una
regularizaci\'on {\it point-splitting} invariante de gauge. 

En el segundo paso hemos calculado expl\'{\i}citamente el espectro de
$(i\not \! \partial +P_0 )_{\ka }$, y mostrado que la funci\'on
$\zeta_{(i\not \partial +P_0 )_{\ka }}$ es anal\'{\i}tica en $s=0$. 

Finalmente, hemos verificado que nuestro resultado es consistente con el
teorema del \'{\i}ndice en variedades con borde. 

\bigskip

En\cite{a1,a2} hemos establecido una relaci\'on entre el
$\zeta$-determinante y la funci\'on de Green del problema, basada en las
potencias complejas de Seeley para el caso de condiciones de contorno
locales.  La aplicaci\'on de la expresi\'on all\'{\i} encontrada al
ejemplo considerado conduce al mismo resultado. Pero en este caso (con
condiciones globales), su relaci\'on con la funci\'on $\zeta$ del operador
no est\'a garantizada.


\end{document}